\documentclass[referee]{aa}
\usepackage{graphicx}
\begin{document}
\title{On the collisional depolarization and transfer rates of spectral lines by atomic hydrogen. III: application to $f$-states of neutral atoms.}

   \author{M. Derouich
          \inst{1},
           S. Sahal-Br\'echot
          \inst{1},
             and
          P. S. Barklem
          \inst{2}
          }
\titlerunning{Collisional depolarization  and transfer rates.}
\authorrunning{M. Derouich et al}
   \institute{Observatoire de Paris-Meudon, LERMA UMR CNRS 8112, 5, Place Jules Janssen, F-92195 Meudon Cedex, France. 
          \and
Department of Astronomy and Space Physics, Uppsala University, Box 515, S 
751 20 Uppsala, Sweden\\
   \email{Moncef.Derouich@obspm.fr}
             }
   \date{Received 2003 / accepted XXXX}

\abstract{
The theory of collisional depolarization of spectral lines by atomic hydrogen
(Derouich et al. 2003a; Derouich et al. 2003b) is extended to $f$-atomic 
levels $(l$=3). Depolarization rates, polarization and population transfer 
rates are calculated and results are given. Each cross section as a function of the effective quantum number for a relative velocity of $10$ $\textrm{km s}^{-1}$  is given together with an exponent $\lambda$, if it exists, on the assumption that the cross section varies with velocity as $v^{-\lambda}$.  A  general trends of  depolarization rates, of polarization transfer rates and of population transfer rates are given. A discussion of our results is achieved.

}
\maketitle
\section{Introduction}
The observation of the so-called ``second solar spectrum'' (a term first suggested by V.V. Ivanov of  St. Petersburg, Russia; see Stenflo \& Keller \cite{Stenflo1}; Stenflo et al. \cite{Stenflo2}; Stenflo  \cite{Stenflo3}; Gandorfer \cite{Gandorfer1};  Gandorfer \cite{Gandorfer2}), which is the spectrum of the linear polarization observed near the limb, is due to the scattering of the underlying anisotropic radiation. The atomic polarization may be modified by several factors, in particular the magnetic field (Hanle effect), and also the isotropic collisions with the neighboring particles of hydrogen. Therefore the depolarization rates, polarization and population transfer rates by collisions with hydrogen are needed in order to quantitatively interpret the observed polarization in terms of magnetic fields in solar quiet regions. 

In Derouich et al. (\cite{derouich1}) and Derouich et al. (\cite{derouich2}) (hereafter Papers I and II, respectively), a semi-classical theory for calculations of depolarization rates, polarization and population transfer rates has been developed and applied  to $p$ $(l$=1) and  $d$ $(l$=2) atomic states. In the present paper we extend this theory to  $f$-atomic levels  $(l$=3).  This paper presents the first calculations of the depolarization  and the collisional transfer rates for $f$-atomic states.

Our semi-classical theory is not specific for a given atom and its application is possible even to heavy atoms (Ti, Fe, ...), for which there are no available depolarization rates and transfer of polarization and population rates data. The extension of this method  gives possibility to calculate 
depolarization and collisional transfer rates of $p$ $(l$=1), $d$ $(l$=2) and $f$ $(l$=3) atomic levels. It should now be
possible to rapidly obtain the large amount of data needed for the interpretation of the second solar spectrum. Using our method, general trends of all rates for $p$ $(l$=1), $d$ $(l$=2) and $f$ $(l$=3) atomic levels with orbital angular momentum quantum number l are able to be discussed for the first time.  
\section{Method}
The method used to determine depolarization rates, transfer of polarization and population rates is the same as the one previously introduced in Papers I and II. We denote as $\displaystyle D^k(n  l J, T)$  the collisional 
depolarization  rate for the statistical tensor of rank $k$. Each 
level of total angular momentum $J$ relaxes with $2J+1$ independent 
depolarization  rates. In particular $D^0(nl J, T)$ is the destruction rate of
 population, which is zero since elastic collisions do not alter the population of an atomic level $(nl J)$, $D^1(nl J, T)$ is the destruction rate of orientation 
(circular polarization) and $D^2(nl  J, T)$ is the destruction rate of 
alignment (linear  polarization) which is of interest in the understanding of the second solar 
spectrum. If the quenching must be taken into account, $\displaystyle D^k(n  l J \to n  l J', T)$ corresponds to collisional transfer of population $(k=0)$, orientation $(k=1)$ and alignment $(k=2)$ (Paper II).   Since  potentials are computed in the rotating frame, which is obtained from  the fixed laboratory frame by means of the geometrical rotation $R$ $(\beta, \frac{\pi}{2},\frac{\pi}{2})$, the interaction potential matrix is 
diagonal (see, for example, Paper I).  The extension of our calculations to $f$-atomic levels $(l$=3) requires the determination of seven RSU potentials  $V_{eff,m}$ $(-3\leq m \leq 3)$. For more details we refer to Paper I and Paper II and to the ABO papers (Anstee \cite{Anstee2}; Anstee \& O'Mara \cite{Anstee1}, \cite{Anstee3}; Anstee, O'Mara \& Ross \cite{Anstee4}; Barklem \cite{Barklem3}; Barklem \& O'Mara \cite{Barklem1}; Barklem, O'Mara \& Ross \cite{Barklem2}). The total wave function  $|\psi \rangle$ of the system is developed over the basis formed by the eigenvectors $ | M_l \rangle$: 
\begin{eqnarray} \label{eq7}
\big | \psi (t) \rangle = \displaystyle \sum_{M_l} a_{M_l}(t)   \textrm{e}^{-\textrm{i}{E^0_{M_l} t}} \big | M_l \rangle 
\end{eqnarray}
where $\displaystyle E^0_{M_l}$ is the energy eigenvalues of the isolated atoms. For $f$-states, seven semi-classical coupled linear differential  equations describing the evolution of the manifold of states are obtained by writing the time-dependent Schr\"odinger equation (see Barklem et al. \cite{Barklem2}).  Having the RSU potentials $V_{eff,m}$, after integration of these coupled  equations over an entire collision, we obtain the $a_{M_l}(t)$ coefficients and then the transition $T$-matrix elements (Paper I; Paper II). All rates are obtained after integrations,  over impact parameters and velocities, of transitions probabilities given by equation (39) in Paper I and equation (11) in Paper II. 
\section{Results}
As  for $p$ and $d$ atomic states calculations, in  most cases, the behaviour of the cross sections with the relative velocity $v$ obeys a power law of the form: 
\begin{eqnarray} \label{eq12}
\sigma^k(n 3 J \to n 3 J', v) (J=J'\; \textrm{and}\; J\ne J')= 
\displaystyle \sigma^k(n 3 J \to n 3 J', v_0)(\frac{v}{v_0})^{-\lambda^k(n 3 J \to n 3 J')} ,
\end{eqnarray}
where ${v_0}$ is a typical velocity where the cross section is calculated ($10$ km s$^{-1}$).   Tables \ref{table1} and \ref{table2} give respectively  variation of cross sections with the   effective principal quantum number $n^*$ and the corresponding velocity exponents. Cross sections for other velocities can 
be obtained  from Tables \ref{table1} and \ref{table2} using equation (\ref{eq12}). Tables \ref{table1} and \ref{table2}  can be interpolated for an appropriate $n^*$  corresponding to a given observed line in order to obtain the needed  rates   (Paper I; Paper II).  For cross sections obeying  equation (\ref{eq12}),  the collisional depolarization and   transfer rates   can be expressed by equation (13) of Paper II. Sometimes, especially for the  alignment transfer  calculations, such behaviour was not obeyed (the cross sections showed oscillations with relative velocities) and so   $\displaystyle \lambda^k(n 3 J \to n 3 J')$  is not reported (Table \ref{table2}). We have to calculate directly the cross sections for each velocity. The collisional depolarization and  transfer rates $D^k(n l J \to n l J', T)$ $(J=J' \; \textrm{and} \; J\ne J')$ follow from  numerical integration over  the velocities. 

\begin{figure}[htbp]
\begin{center}
\includegraphics[width=8 cm]{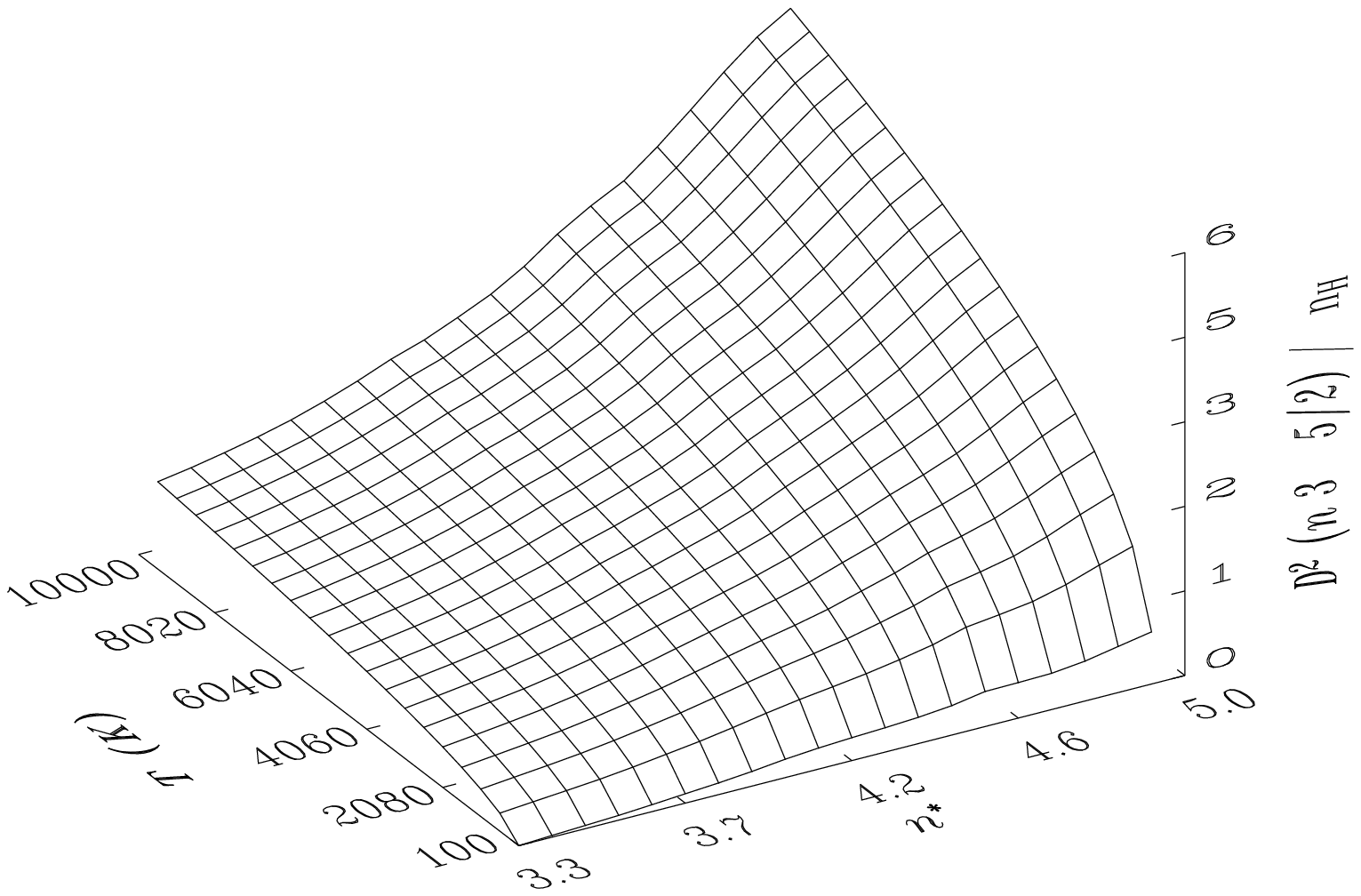}
\includegraphics[width=8 cm]{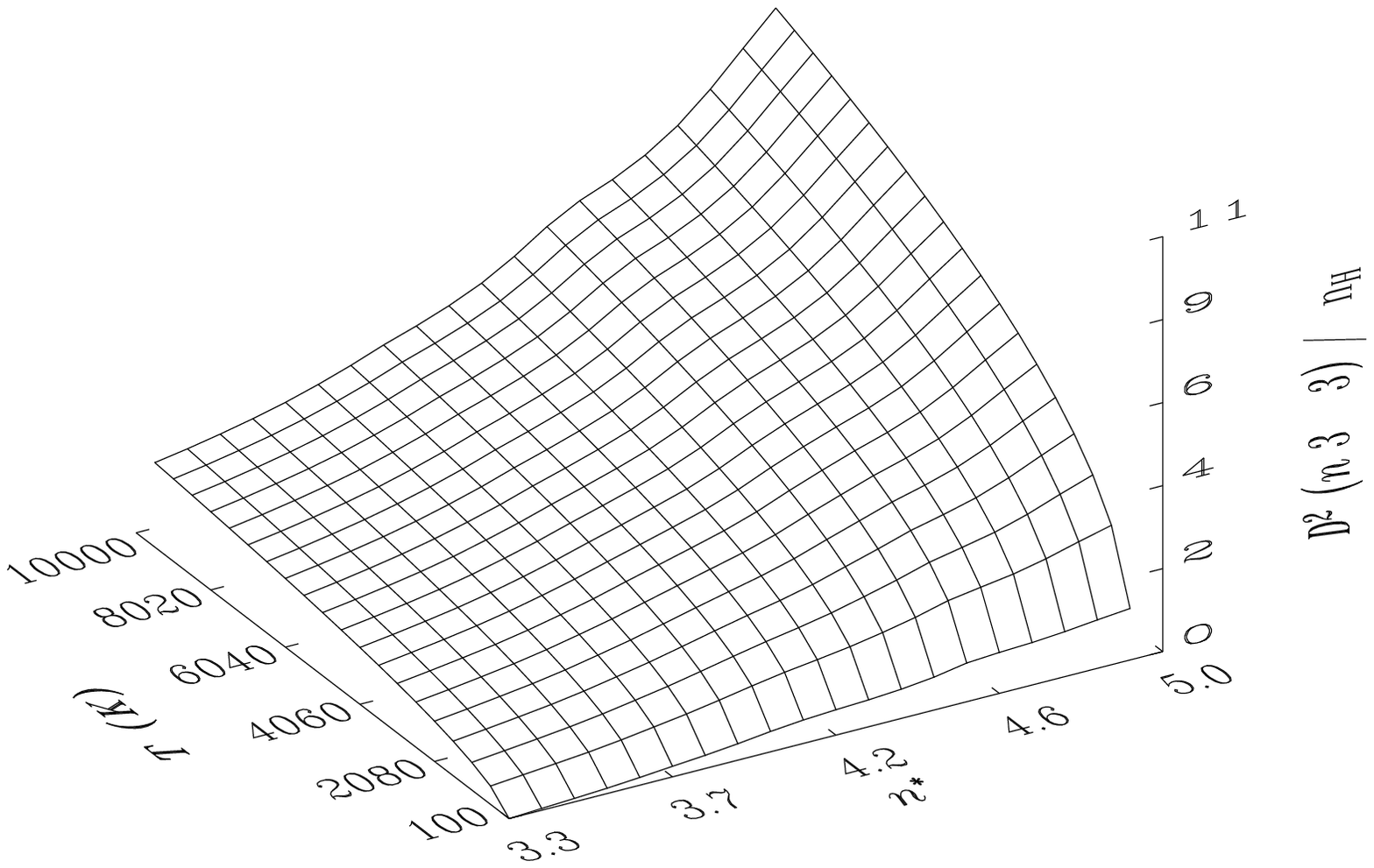}
\includegraphics[width=8 cm]{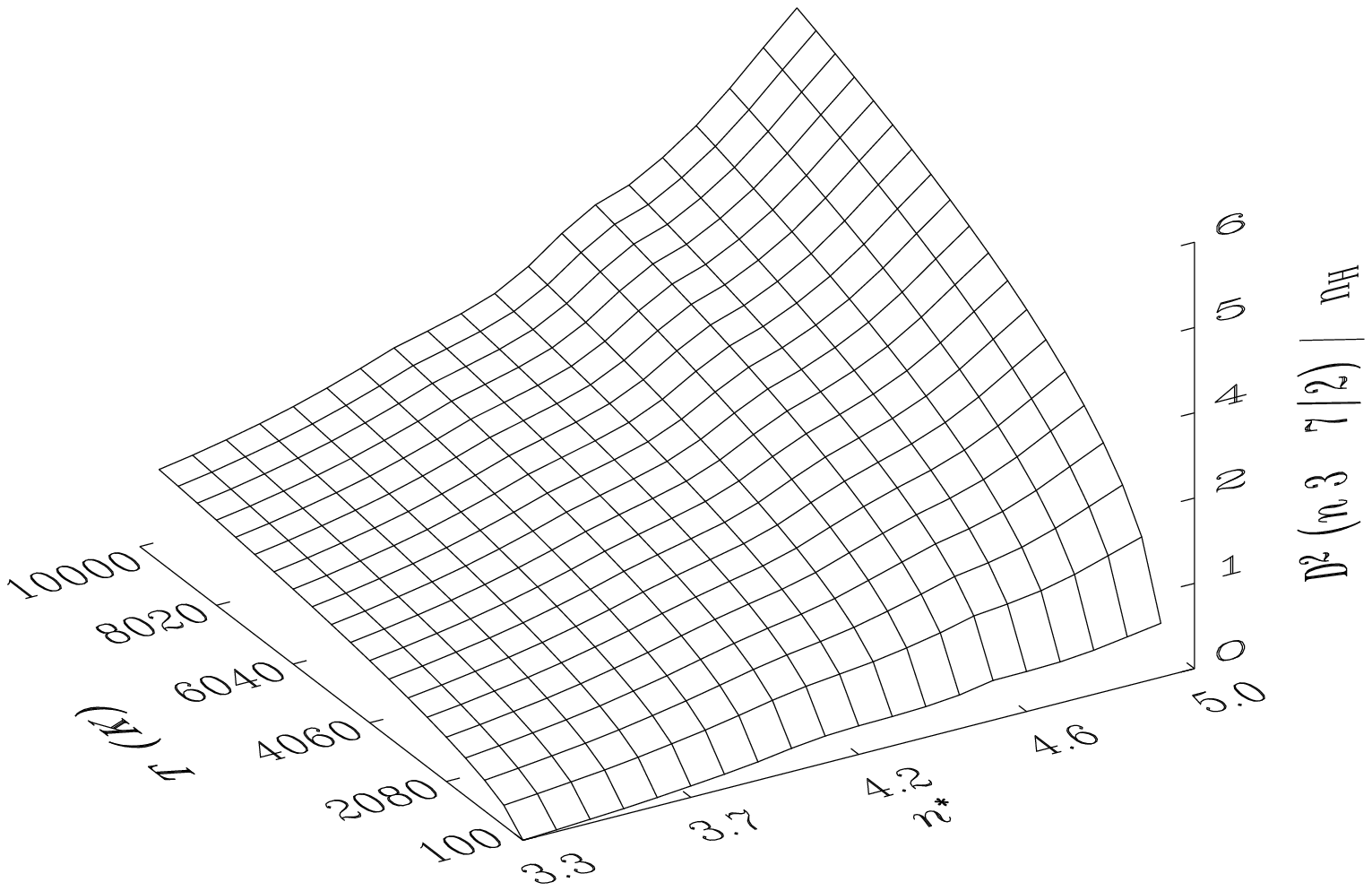}
\end{center}
\caption{Depolarization rates per unit H-atom density as a function of temperature $T$ and $n^*$. For $l=3$, each figure:  $S=\frac{1}{2}$ and $J=\frac{5}{2}$; $S=0$ and $J=3$; $S=\frac{1}{2}$ and $J=\frac{7}{2}$. Depolarization rates are given in $10^{-14}$ $\textrm{rad.} \;  \textrm{m}^3 \; \textrm{s}^{-1}$.}
\label{depolarizationratefunction}
\end{figure}

\begin{figure}[htbp]
\begin{center}
\includegraphics[width=8 cm]{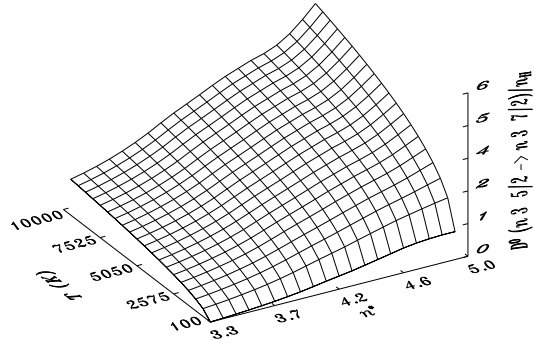}
\end{center}
\caption{Population transfer rate per unit H-atom density ($k$=0) as a function of temperature $T$ and $n^*$. $l=3$, $S=\frac{1}{2}$, $J=\frac{5}{2}$ and $J'=\frac{7}{2}$. Population transfer rate is given in $10^{-14}$ $\textrm{rad.} \;  \textrm{m}^3 \; \textrm{s}^{-1}$.}
\label{transferfunction0}
\end{figure}

\begin{figure}[htbp]
\begin{center}
\includegraphics[width=8 cm]{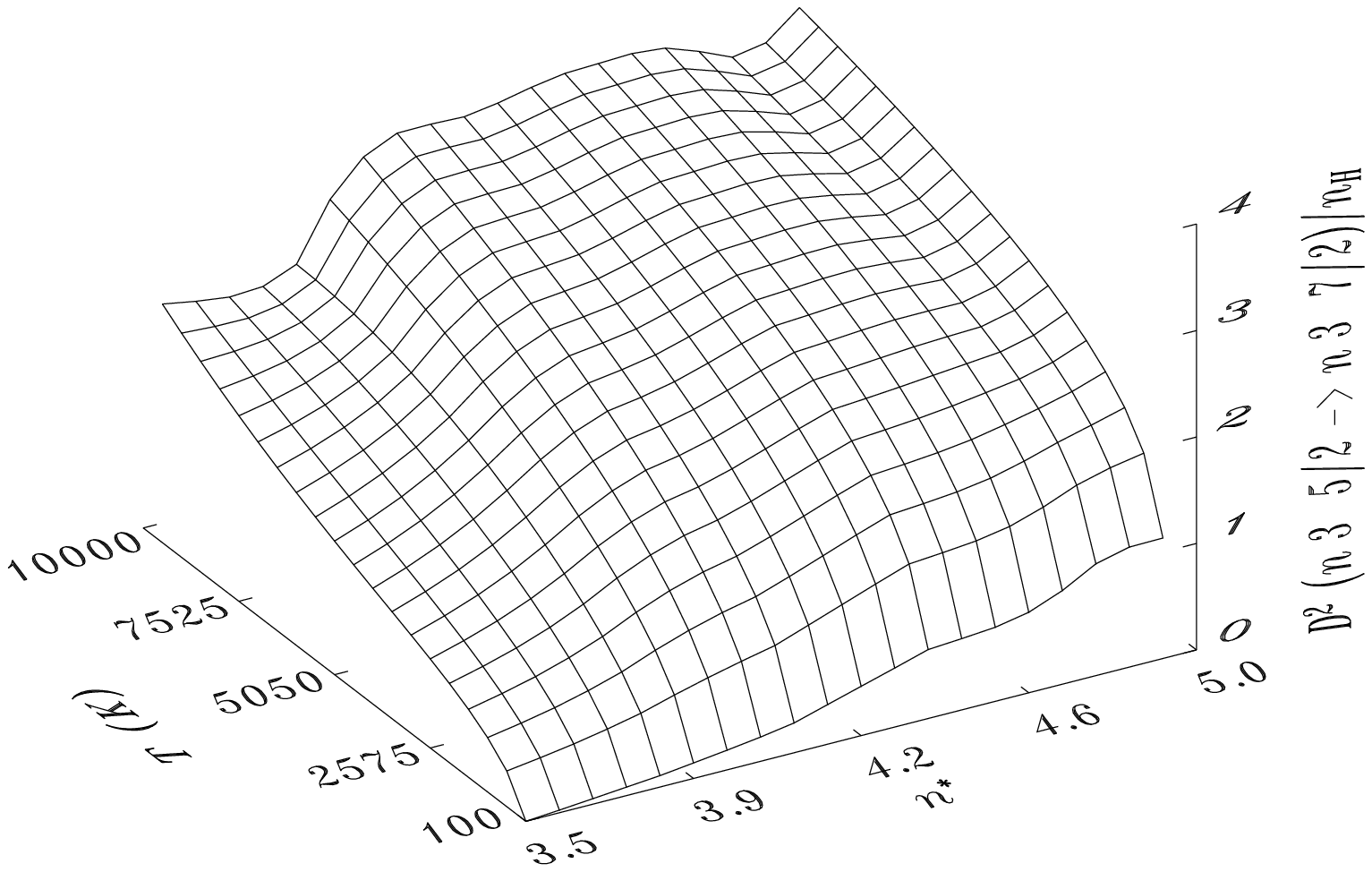}
\end{center}
\caption{Linear polarization transfer rate per unit H-atom density ($k$=2) as a function of temperature $T$ and $n^*$.  $l=3$, $S=\frac{1}{2}$, $J=\frac{5}{2}$ and $J'=\frac{7}{2}$. Linear polarization transfer rate is given in $10^{-15}$ $\textrm{rad.} \;  \textrm{m}^3 \; \textrm{s}^{-1}$.}
\label{transferfunction2}
\end{figure}

Figure \ref{depolarizationratefunction} shows the alignment depolarization rates ($k=2$) as a function of the local temperature $T$ and $n^*$ for $l=3$. The population transfer rates ($k=0$) and the linear polarization transfer rates ($k=2$) as a function of $T$ and $n^*$ are displayed in Figs \ref{transferfunction0} and \ref{transferfunction2}. All these rates increase with the temperature. For a temperature $\displaystyle T 
\leq $ 10000 K,  the destruction rate of alignment  $\displaystyle D^2(n \; 3
 \; 5/2)/n_H \leq 7  \times 10^{-14}$ $\textrm{rad.} \;  \textrm{m}^3 \; \textrm{s}
^{-1}$, $\displaystyle D^2(n \; 3 \; 3)/n_H \leq 12 \times 10^{-14}$ $\textrm{rad.}
 \;  \textrm{m}^3 \; \textrm{s}^{-1}$  and $\displaystyle D^2(n \; 3 \; 7/2)/ 
n_H \leq 7  \times 10^{-14}$ $\textrm{rad.} \;  \textrm{m}^3 \; \textrm{s}^{-1}$. The population transfer rate $\displaystyle D^0(n \; 3 \; 5/2  \to n \; 3 \; 
7/2)/n_H \leq 7  \times 10^{-14}$ $\textrm{rad.} \;  \textrm{m}^3 \; \textrm{s}^
{-1}$ and the linear  polarization transfer rate $\displaystyle D^2(n \; 3 \; 
5/2  \to n \; 3 \; 7/2)/n_H \leq 5  \times 10^{-15}$ $\textrm{rad.} \;  \textrm{m}^
3 \; \textrm{s}^{-1}$. These numerical values are given for  $\displaystyle 
n^* \leq 5$ which include most of the lines of  interest. 
\begin{table}
\begin{center}
\begin{tabular}{|l|c|c|c|c|c|r|}
\hline
$n^*$ & $\sigma^2(n 3 \frac{5}{2})$ &$\sigma^2(n 3 3)$& $\sigma^2(n 3 \frac{7}{2})$ & $\sigma^0(n 3 \frac{5}{2} \to n 3 \frac{7}{2})$ & $\sigma^2 (n 3 \frac{5}{2} \to n 3 \frac{7}{2})$\\
\hline
$3.3$ & 426& 782& 491& 478&40	 \\
\hline
$3.4$ & 451& 831& 521& 518&44	 \\
\hline 
$3.5$ &489& 892& 561&565 &51  	\\
\hline 
$3.6$ &	535& 962& 603&616 &58	\\
\hline 
$3.7$&	584& 1055& 654&684 &67	\\
\hline
$3.8$&	637& 1151& 708&768 & 81	\\
\hline 
$3.9$&695& 1249	&765 &865 &96 \\
\hline
$4$&759&1365	 &824&962 &101 \\
\hline 
$4.1$ &835 &1467 &875	&1031 &110 \\
\hline 
$4.2$ &890  & 1572&899	&1119 &108\\
\hline 
$4.3$ &1024& 1777&1051	&1188 &114 \\
\hline 
$4.4$ &1077&1908	&1097	&1297 &121 \\
\hline 
$4.5$ &1236&2166	&1247	&1418 &121\\
\hline 
$4.6$ &	1249&2279& 1306	&1508 &116\\
\hline 
$4.7$ &1420& 2411&1432	&1503&125\\
\hline 
$4.8$ &1485& 2515&1485	&1529 &118\\
\hline 
$4.9$ &1521&2587	&1504	&1630&132\\ 
\hline 
$5$ &1928& 3135	&1845	&1786&130\\ 
\hline 
\end{tabular}
\end{center}
\caption{Variation of the  cross sections, for the relative velocity  of $10 \;\textrm{km} \; \textrm{s}^{-1}$,  with the effective principal number. Cross sections are in atomic units.}
\label{table1}
\end{table} 
\begin{table}
\begin{center}
\begin{tabular}{|l|c|c|c|c|r|}
\hline
$n^*$ & $\lambda^2(n 3 \frac{5}{2})$ & $\lambda^2(n 3 3)$& $\lambda^2(n 3 \frac{7}{2})$ & $\lambda^0(n 3 \frac{5}{2} \to n 3 \frac{7}{2})$ & $\lambda^2 (n 3 \frac{5}{2} \to n 3 \frac{7}{2})$\\
\hline
$3.3$ &0.228 & 0.277& 0.301 &0.280&-	\\
\hline 
$3.4$ &0.249 & 0.295& 0.315 &0.283 & - 	\\
\hline 
$3.5$ &0.252	&0.303& 0.325 &0.288& -	\\
\hline 
$3.6$&0.260&0.311& 0.331 &0.289	& -	\\
\hline
$3.7$&0.275& 0.329& 0.346&0.289	&-	\\
\hline 
$3.8$&0.275 &0.334&0.354	&0.296	&- \\
\hline
$3.9$&0.282&0.342&0.367 &0.301&0.103 \\
\hline 
$4$ &0.306&0.363&0.402	&0.320&0.117 \\
\hline 
$4.1$ &0.298&0.362	&0.412	&0.344&0.176\\
\hline 
$4.2$ &0.278&0.336&0.384	&0.351&0.267 \\
\hline 
$4.3$ &0.240&0.306	&0.333	&0.370&0.328 \\
\hline 
$4.4$ & 0.235&0.304	&0.319	&0.394&0.384\\
\hline 
$4.5$ &0.262&0.333	&0.349 &0.415&0.396\\
\hline 
$4.6$ &	0.215&0.317 &0.311	&0.440&0.411\\
\hline 
$4.7$ &	0.181&0.289 &0.284 &0.435&0.497\\
\hline 
$4.8$ &0.192&0.268	&0.269	&0.409&0.600\\ 
\hline 
$4.9$ &0.218&0.248	&0.268	&0.363&0.587\\ 
\hline 
$5$ &0.222&0.224	&0.242	&0.331&0.500\\ 
\hline  
\end{tabular}
\end{center}
\caption{Velocity exponents $\displaystyle \lambda^k(n l J \to n l J') (J=J' \textrm{and} J\ne J')$ corresponding to the cross sections of Table \ref{table1}.}
\label{table2}
\end{table}
\section{General trends}
For a given effective quantum number $n^*$, and for the cases $l=1$, $l=2$, and $l=3$, destruction rates of alignment are such that $\displaystyle D^2(n \; 3 \; 3, T)$ $<$ $\displaystyle D^2(n \; 2 \; 2, T)$ $<$ $\displaystyle D^2(n \; 1 \; 1, T)$. A similar result has been obtained for the broadening of spectral lines. In fact, Barklem et al. (\cite{Barklem2}) have previously shown that, also for a given $n^*$, lines with upper $p$-states ($l=1$) are more broadened  than lines with upper $d$-states ($l=2$), and similarly lines with upper $d$-states are more broadened than lines with upper $f$-states ($l=3$). This effect is similar to that first seen observationally in the solar spectrum by Carter (\cite{carter1}). In general,  when the orbital angular momentum quantum number $l$ increases the  depolarization rates and transfer of polarization and population rates decrease for a given value of the energy of the state of the valence electron $\displaystyle E_{n l }$. $\displaystyle E_{n l }$ is related to $ n^*$ by $ n^*=[2 (E_\infty - E_{n l })]^{-1/2}$, $\displaystyle E_\infty$ being the binding energy of the ground state.

For  $f$-states, when $J=7/2$ and $J'=5/2$ we have: $\displaystyle 
D^0(n 3 7/2 \to n 3 5/2, T)$ $>$ $\displaystyle D^1(n 3 7/2 \to n 3 5/2, T)$ $>$ $\displaystyle D^2(n 3 7/2 \to n 3 5/2, T)$ $>$ 
$\displaystyle D^4(n 3 7/2 \to n 3 5/2, T)$ $>$ $\displaystyle D^3(n 3 7/2 \to n 3 5/2, T)$ $>$ $\displaystyle D^5(n 3 7/2 \to n 3 5/2, T)$. We recall that $D^k(n \; l \; J \to n \; l \; J', T)$   is a linear combination of  $\zeta (n l J M_J \to n l J' M'_J, T)$ (equation (3) in Paper II). The population transfer rate is the greater transfer rate because for $k = 0$ the coefficients of this linear combination are  positive.  These coefficients are constant and equal to $1/\sqrt{(2J+1)(2J'+1)}$ which leads to a  $\displaystyle D^0(n l J \to n l J', T)$ which is proportional to $\zeta (n l J \to n l J', T)$ (equation (5) in Paper II).  However, the sign of the coefficients of the linear combination  for transfer rates of rank $k\geq1$  is sometimes positive and sometimes negative. For example, these  coefficients have the sign of $M_J \times M'_J$ for orientation transfer rates $(k=1)$ and the sign of $(3M^2_J - J(J+1)) \times (3M'^2_J - J'(J'+1))$ for alignment transfer rates. The other coefficients of the linear combination for $k > 2$ may  be obtained on request from the authors. We conclude that, for $k \neq 0$, the collisional transfer rates may be positive or negative as a function of transition probabilities between the Zeeman sublevels which depend on $n^*$. The depolarization rates are usually positive.

All rates were found to increase with temperature $T$. The functional form $\displaystyle D(T) = A T ^{(1-\lambda)/2}$ may be accurately fitted to the depolarization rates and the population transfer rates (Paper I; Paper II). However, sometimes  the collisional transfer rates with $k \neq 0$, cannot be fitted by the power-law $A T ^{(1-\lambda)/2}$ and so ${\lambda}$ is not reported (Table \ref{table2} in the present paper and Table 2 in Paper II). This is due to the fact that these  collisional transfer rates  are the sum of incoherent contributions from the states $|n l J M_J \rangle$ and $|n l J' M'_J \rangle$.  We notice  that  the above remarks are valid also for $p$ and $d$-atomic states.   
\section{Discussion}
Unfortunately,  there is neither  experimental nor quantum chemistry 
depolarization  and collisional transfer rates for $f$-states for comparison. 
We expect that the main differences between the  RSU potentials and those from quantum chemistry, which are considered as more realistic, occur at the short-range interactions. We have verified that these close collisions do not influence the computed depolarization  and collisional transfer rates for $f$-states. The decisive contribution to the  depolarization  and collisional transfer rates calculations occurs at intermediate-range interactions. In Paper I, which is concerned with $p$-states, comparison with  quantum chemistry results in  Kerkeni  (\cite{Kerkeni2}) gives   depolarization rates in agreement to  better than  20 \%. Extrapolating our results obtained for  $p$ and $d$ states (Paper I; Paper II), we expect  a rather good agreement (relative difference less than 20 \% at solar temperatures) between our rates obtained for $f$-states and a full quantum mechanical treatment.
\section{Conclusion}
This paper is a continuation of a series concerned with the theoretical calculations of the  depolarization  and collisional transfer rates. Thanks to the extension to $f$-atomic states ($l=3$), we are able to  obtain the first general conclusions concerning trends of all rates as a function of orbital angular momentum quantum number $l$. An extrapolation for $l$ $>$ 3 would be useful for a more complete interpretation of the ``second solar spectrum''. This work is in progress. An extension of our theory to the case of ions   will be the subject of further papers. 

\end{document}